\documentclass{article}
\usepackage{spconf,amsmath,graphicx,amsfonts,amssymb,stfloats,booktabs,multirow,bbding,pifont,url,color,xcolor,makecell,cite,bm}
\usepackage[colorlinks,linkcolor=black,anchorcolor=black,citecolor=black,urlcolor=black]{hyperref}

\def\L{{\cal L}}

\newcommand{\cmark}{\ding{51}}%
\newcommand{\xmark}{\ding{55}}%

\title{Unifying Speech Enhancement and Separation with Gradient Modulation for End-to-End Noise-Robust Speech Separation}
\name{Yuchen Hu$^1$, Chen Chen$^1$, Heqing Zou$^1$, Xionghu Zhong$^2$, Eng Siong Chng$^1$}
\address{$^1$School of Computer Science and Engineering, Nanyang Technological University, Singapore \\ $^2$College of Computer Science and Electronic Engineering, Hunan University, China}

\begin{document}
\ninept
\maketitle
\begin{abstract}
Recent studies in neural network-based monaural speech separation (SS) have achieved a remarkable success thanks to increasing ability of long sequence modeling.
However, they would degrade significantly when put under realistic noisy conditions, as the background noise could be mistaken for speaker's speech and thus interfere with the separated sources.
To alleviate this problem, we propose a novel network to unify speech enhancement and separation with gradient modulation to improve noise-robustness.
Specifically, we first build a unified network by combining speech enhancement (SE) and separation modules, with multi-task learning for optimization, where SE is supervised by parallel clean mixture to reduce noise for downstream speech separation.
Furthermore, in order to avoid suppressing valid speaker information when reducing noise, we propose a gradient modulation (GM) strategy to harmonize the SE and SS tasks from optimization view.
Experimental results show that our approach achieves the state-of-the-art on large-scale Libri2Mix- and Libri3Mix-noisy datasets, with SI-SNRi results of 16.0 dB and 15.8 dB respectively.
Our code is available at GitHub\footnote{\scriptsize\url{https://github.com/YUCHEN005/Unified-Enhance-Separation}}.
\end{abstract}

% Recent studies in neural network-based monaural speech separation (SS) have achieved a remarkable success thanks to increasing ability of long sequence modeling. However, they would degrade significantly when put under realistic noisy conditions, as the background noise could be mistaken for speaker's speech and thus interfere with the separated sources. To alleviate this problem, we propose a novel network to unify speech enhancement and separation with gradient modulation to improve noise-robustness. Specifically, we first build a unified network by combining speech enhancement (SE) and separation modules, with multi-task learning for optimization, where SE is supervised by parallel clean mixture to reduce noise for downstream speech separation. Furthermore, in order to avoid suppressing valid speaker information when reducing noise, we propose a gradient modulation (GM) strategy to harmonize the SE and SS tasks from optimization view. Experimental results show that our approach achieves the state-of-the-art on large-scale Libri2Mix- and Libri3Mix-noisy datasets, with SI-SNRi results of 16.0 dB and 15.8 dB respectively. Our code is available at GitHub.


\vspace{-0.28cm}
\section{Introduction}
\label{sec:intro}
\vspace{-0.22cm}
Recent progress in neural network-based monaural speech separation (SS) has achieved a remarkable success in time-domain methods~\cite{luo2019conv, luo2020dual, subakan2021attention, subakan2022using, zeghidour2021wavesplit}, thanks to the increasing long sequence modeling ability of dilated CNN, RNN and Transformer~\cite{vaswani2017attention}.
As a result, the time-domain methods outperform conventional time-frequency domain methods and achieve state-of-the-art on various benchmarks.

However, their performance would degrade significantly when put under the real-world noisy conditions.
The reason could be that some background noise is mistaken for speaker's speech and thus interferes with the separated sources~\cite{chen2016noise}, just like we humans also get confused under noisy mixture scenes.
Current studies on end-to-end noise-robust speech separation are quite limited.

Speech enhancement (SE)~\cite{wang2020complex} has been proved effective in reducing noise from the noisy speech to improve speech quality for many downstream tasks, \textit{e.g.}, automatic speech recognition~\cite{liu2021ustc,chen2022self,chen2022noise,zhu2022noise,zhu2022joint,zhu2022robust}, with multi-task learning to make full use of the clean supervision information~\cite{pandey2021dual}.
Nevertheless, such joint network could also bring another over-suppression problem~\cite{hu2022interactive,hu2022dual} where SE suppresses some important speech information together with the background noise, resulting in sub-optimal performance for downstream tasks.
From the optimization view, there could exist conflicts between the gradients of SE task and downstream tasks, which would hinder the multi-task learning and degrade the downstream task performance.

In this paper, we propose a novel network to unify speech enhancement and separation with gradient modulation for end-to-end noise-robust speech separation.
Specifically, we first build a unified network by combining speech enhancement and separation modules, where the front-end SE serves to reduce noise for back-end speech separation.
Multi-task learning strategy is employed to make full use of the supervision information in parallel clean mixture, which is available in most benchmark datasets since they usually create noisy mixtures by simulation.
Furthermore, in order to avoid suppressing valid speaker information when reducing noise, we propose a gradient modulation (GM) strategy to harmonize the SE and SS tasks from optimization view.
Experimental results indicate that our proposed approach improves the noise-robustness of speech separation model and significantly outperforms the previous state-of-the-arts.
To the best of our knowledge, this is the first exploration to unify speech enhancement and separation with harmonized multi-task learning.

\begin{figure*}[t]
  \centering
  \includegraphics[width=0.93\textwidth]{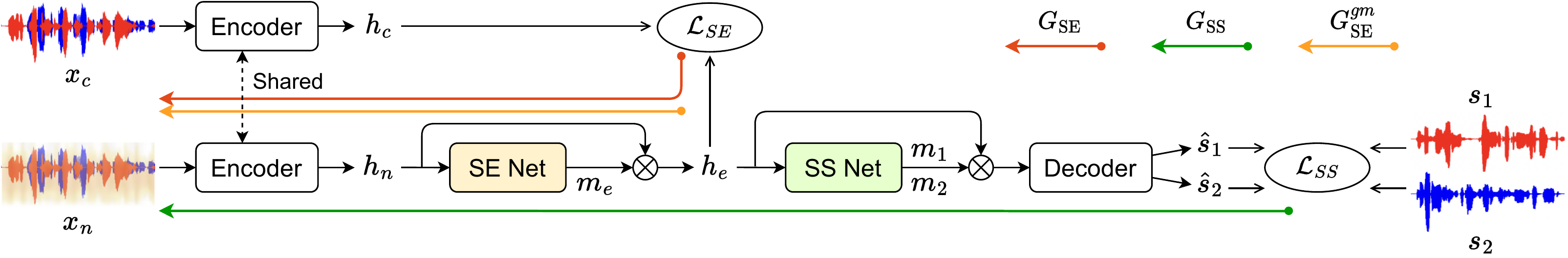}
  \vspace{-0.35cm}
  \caption{The overall architecture of our proposed unified network, which consists of encoder, SE network, SS network and decoder.
  The $x_n$ denotes noisy mixture, $x_c$ denotes parallel clean mixture, $s_1$ and $s_2$ denote the target sources.
  The $\bm{G}$ denotes gradient.}\label{fig1}
  \vspace{-0.3cm}
\end{figure*}

\vspace{-0.15cm}
\section{Proposed Method}
\label{sec:proposed_method}
\vspace{-0.25cm}
\subsection{System Overview}
\label{ssec:overview}
\vspace{-0.1cm}
In this work, we first build a unified network in Figure~\ref{fig1} with multi-task learning for optimization, where the SE module is supervised by parallel clean mixture to reduce noise for back-end speech separation.
However, we observe that apart from noise, the SE module can also suppress some valid speaker information, \textit{i.e.}, over-suppression problem~\cite{hu2022interactive}, which may degrade the downstream SS performance.
From the optimization view, there exists some conflicts between the gradients of SE and SS tasks, which would hinder the multi-task learning and finally lead to sub-optimal SS performance.
To this end, we propose a gradient modulation strategy to harmonize the two task gradients and thus alleviate the over-suppression problem.

\vspace{-0.3cm}
\subsection{Unified Network}
\label{ssec:unified_net}
\vspace{-0.15cm}

\subsubsection{Architecture}
\label{sssec:arch}

\vspace{-0.15cm}
Inspired by the popular mask-learning framework~\cite{luo2019conv, luo2020dual, subakan2021attention, subakan2022using}, we design a unified network that consists of an encoder, a speech enhancement (SE) network, a speech separation (SS) network and a decoder, as shown in Figure~\ref{fig1}.
The encoder first learns deep representations of the input mixtures.
The SE network then learns a mask to filter out background noise, followed by SS network to predict multiple masks to separate different sources in the mixture.
Finally, the decoder reconstructs the separated speech in the time domain.

\vspace{0.1cm}
\noindent\textbf{Encoder.} The encoder takes in the time-domain noisy mixture $x_n \in \mathbb R^T$ and learns a STFT-like representation $h_n \in \mathbb R^{F \times T'}$ using a convolutional layer.
The same process is done for the parallel clean mixture $x_c \in \mathbb R^T$, generating a clean representation $h_c \in \mathbb R^{F \times T'}$ to supervise speech enhancement task:
\vspace{-0.05cm}
\begin{equation}
\label{eq1}
\begin{split}
    h_n &= \text{ReLU}(\text{Conv1d}(x_n)), \\
    h_c &= \text{ReLU}(\text{Conv1d}(x_c)),
\end{split}
\end{equation}

\noindent\textbf{Speech Enhancement (SE) Network.} The SE network takes in the noisy mixture representation $h_n$ and learns a mask $m_e$ to filter out background noise, resulting in enhanced mixture representation $h_e$:
\begin{equation}
\label{eq2}
\begin{split}
    m_e &= \text{SE-Net}(h_n), \\
    h_e &= m_e \cdot h_n,
\end{split}
\end{equation}

\noindent where the SE network follows the same architecture as SS network described in Figure~\ref{fig3}.
In particular, SE network is a special case of the SS network where the number of separated sources is 1.

The generated $h_e$ is employed to calculate speech enhancement loss by compared to the clean mixture representation $h_c$, and then it will be sent into the SS network for source separation.

\vspace{0.1cm}
\noindent\textbf{Speech Separation (SS) Network.} Figure~\ref{fig3} illustrates the architecture of SS network, which follows prior works like Dual-Path RNN~\cite{luo2020dual} and SepFormer~\cite{subakan2021attention}.
It takes in the enhanced mixture representation $h_e$ and predicts a mask $m_k \in \{m_1, m_2, \dots,m_{C}\}$ for each of $C$ sources in the mixture.
\begin{equation}
\label{eq3}
\begin{split}
    m_k &= \text{SS-Net}(h_e), \\
\end{split}
\end{equation}
\vspace{-0.4cm}

As shown in Figure~\ref{fig3}(a), the SS network first sends the input representation $h_e$ into layer normalization and a linear layer with output dimension $F$.
Then, it creates overlapping chunks of size $K$ by chopping up $h_e$ along the time axis with 50\% overlap between neighbors, resulting in an output $h'_e \in \mathbb R^{F \times K \times S}$, where $K$ is chunk length and $S$ is the resulted number of chunks. 

The representation $h'_e$ then feeds the sequence modeling block as illustrated in Figure~\ref{fig3}(b), where we employ Dual-Path RNN block~\cite{luo2020dual} or SepFormer block~\cite{subakan2021attention} with dual-path structure to learn both local and global contexts for long sequence modeling.

After that, the output $h''_e \in \mathbb R^{F \times K \times S}$ is processed by parametric ReLU (PReLU) activation~\cite{he2015delving} and a linear layer. 
The output is denoted as $h'''_e \in \mathbb R^{(C \times F) \times K \times S}$, where $C$ is the number of sources.
Following this, the overlap-add scheme described in~\cite{luo2020dual} is employed to obtain $h''''_e\in \mathbb R^{(C \times F) \times T'}$.
This representation finally goes through two feed-forward layers and a ReLU~\cite{glorot2011deep} activation to generate source masks, which are divided along the channel dimension for each source $k \in \{1, 2, \dots, C\}$.

\vspace{0.1cm}
\noindent\textbf{Decoder.} The decoder takes in the element-wise multiplication of source mask $m_k$ and enhanced mixture representation $h_e$, and reconstruct the separated speech in time-domain using a transposed convolution layer with same kernel size and stride as the encoder.
The transformation is expressed as: 
\vspace{-0.05cm}
\begin{equation}
\label{eq4}
\begin{split}
    \hat {s}_k &= \text{TransposeConv1d}(m_k * h_e),
\end{split}
\vspace{-0.2cm}
\end{equation}
where $\hat{s}_k \in \mathbb R^T$ is the separated speech for source $k$.

\vspace{-0.1cm}
\subsubsection{Multi-task Learning}
\label{sssec:multitask}
\vspace{-0.1cm}
We build two training objectives to optimize the unified network, \textit{i.e.}, SE loss and SS loss.
Firstly, the SE loss is calculated via mean square error (MSE) between enhanced and clean mixture representations, in order to direct the SE network to produce better $h_e$ for separation, making full use of the supervision information in clean mixture:
\vspace{-0.2cm}
\begin{equation}
\label{eq5}
\begin{split}
    \mathcal{L}_\text{SE} &= \frac{1}{F T'} \Vert h_e - h_c\Vert_2^2,
\end{split}
\end{equation}

\vspace{-0.2cm}
\noindent where $\Vert \cdot \Vert_2$ denotes $L_2$ norm, $h_e, h_c \in \mathbb R^{F \times T'}$, $F$ denotes the embedding size and $T'$ denotes the sequence length.

Secondly, the SS loss $\mathcal{L}_\text{SS}$ is calculated using scale-invariant signal-to-noise ratio (SI-SNR)~\cite{le2019sdr} via utterance-level permutation invariant loss~\cite{kolbaek2017multitalker}, following prior works~\cite{luo2019conv, luo2020dual, subakan2021attention}.

The entire system is optimized in an end-to-end manner via multi-task learning strategy, where the overall training objective is formed as: $\mathcal{L} = \lambda_\text{SE} \cdot \mathcal{L}_\text{SE} + \mathcal{L}_\text{SS}$, where $\lambda_\text{SE}$ is a weighting parameter.

\begin{figure}[t]
  \centering
  \vspace{0.1cm}
  \includegraphics[width=0.57\columnwidth]{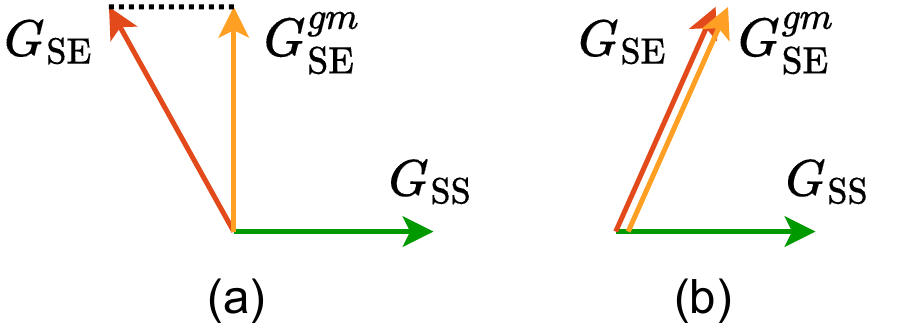}
  \vspace{-0.35cm}
  \caption{Block diagrams of gradient modulation: (a) If $\bm{G}_\text{SE}$ conflicts with $\bm{G}_\text{SS}$ (\textit{i.e.}, the angle between them is larger than $90^\circ$), we set the updated $\bm{G}^{gm}_\text{SE}$ as the projection of $\bm{G}_\text{SE}$ on the normal plane of $\bm{G}_\text{SS}$, (b) If $\bm{G}_\text{SE}$ is aligned with $\bm{G}_\text{SS}$, we set $\bm{G}^{gm}_\text{SE}$ equals to $\bm{G}_\text{SE}$.}\label{fig2}
  \vspace{-0.45cm}
\end{figure}

\vspace{-0.2cm}
\subsection{Gradient Modulation (GM)}
\label{ssec:grad_modulation}
\vspace{-0.1cm}
From the back-propagation view, we denote the SE task gradient as $\bm{G}_\text{SE} = \nabla_v (\lambda_\text{SE}\cdot\mathcal{L}_\text{SE})$, and the SS task gradient as $\bm{G}_\text{SS} = \nabla_v \mathcal{L}_\text{SS}$, where $v$ stands for model parameters.
As shown in Figure~\ref{fig1}, $\bm{G}_\text{SE}$ goes back through SE network and encoder (red arrow), and $\bm{G}_\text{SS}$ passes the entire system (green arrow).
Therefore, the front-end SE network and encoder would be optimized by both gradients, so that the overall gradient can be expressed as follows:
\vspace{-0.1cm}
\begin{equation}
\label{eq6}
\begin{split}
    \bm{G} = \bm{G}_\text{SE} + \bm{G}_\text{SS},
\end{split}
\end{equation}
\vspace{-0.45cm}

However, we have observed some conflicts between the gradients $\bm{G}_\text{SE}$ and $\bm{G}_\text{SS}$, \textit{i.e.}, the angle between them is larger than $90^\circ$.
It indicates that the SE gradient is hindering, instead of assisting in, the optimization of SS task, which is essentially the same as the over-suppression problem described in Section~\ref{ssec:overview}.

To this end, we propose a gradient modulation (GM) strategy to harmonize the two task gradients, as illustrated in Figure~\ref{fig2}:
(1) The angle between $\bm{G}_\text{SE}$ and $\bm{G}_\text{SS}$ is larger than $90^\circ$ (Figure~\ref{fig2}(a)), which means they are conflicting and thus the SE gradient will hinder the optimization of SS task.
In this case, we project $\bm{G}_\text{SE}$ to the normal plane of $\bm{G}_\text{SS}$ to remove the conflict, which avoids increasing the SS loss.
(2) Their angle is smaller than $90^\circ$ (Figure~\ref{fig2}(b)), which means there is no conflict between the two gradients, so that we safely set the updated SE gradient $\bm{G}^{gm}_\text{SE}$ as $\bm{G}_\text{SE}$.
Therefore, our gradient modulation strategy is mathematically formulated as:
\vspace{-0.1cm}
\begin{equation}
\label{eq7}
\bm{G}^{gm}_\text{SE}=
\left\{
\begin{array}{ll}
    \bm{G}_\text{SE} - \frac{\bm{G}_\text{SE} \cdot \bm{G}_\text{SS}}{\Vert \bm{G}_\text{SS} \Vert_2^2} \cdot \bm{G}_\text{SS}, \hspace{0.2cm}& \text{if}\hspace{0.2cm} \bm{G}_\text{SE} \cdot \bm{G}_\text{SS} < 0 \\
    \bm{G}_\text{SE}, \hspace{0.2cm}& \text{otherwise}. 
\end{array}
\right.
\end{equation}

In particular, this strategy is conducted on the gradients of each layer in SE network and encoder, which are flatten to 1-dimensional long vectors in advance and reshaped back after modulation.

As a result, the two task gradients would harmony with each other after modulation and promote the multi-task learning, where the auxiliary SE task could effectively reduce noise to benefit the target SS task while without suppressing valid speaker information.
Finally, the overall gradient is formulated as follows:
\vspace{-0.15cm}
\begin{equation}
\label{eq8}
\begin{split}
    \bm{G}^{gm} = \bm{G}^{gm}_\text{SE} + \bm{G}_\text{SS},
\end{split}
\vspace{0.1cm}
\end{equation}

\begin{figure*}[t]
  \centering
  \includegraphics[width=0.88\textwidth]{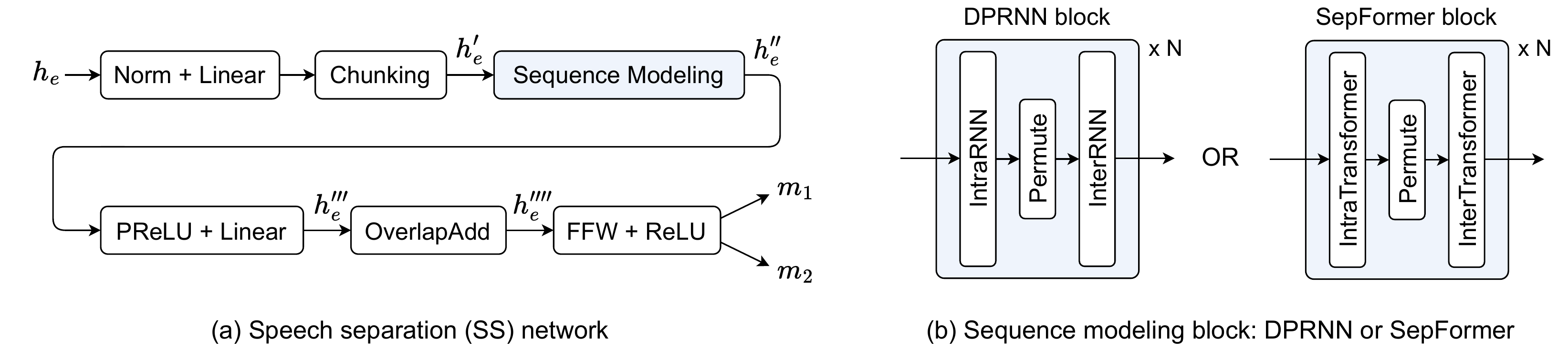}
  \vspace{-0.3cm}
  \caption{Block diagrams of speech separation (SS) network: (a) overall architecture, (b) sequence modeling block.}\label{fig3}
  \vspace{-0.45cm}
\end{figure*}

\vspace{-0.45cm}
\section{Experiments and Results}
\label{sec:exp_result}
\vspace{-0.3cm}

\subsection{Datasets}
\label{ssec:datasets}
\vspace{-0.2cm}
We conduct experiments on the large-scale benchmark Libri2Mix and Libri3Mix~\cite{cosentino2020librimix} datasets\footnote{\scriptsize\url{https://github.com/JorisCos/LibriMix}} (noisy version) to evaluate our proposed approach, where all the waveforms are sampled at 8 kHz.

Libri2Mix is a popular benchmark for speech separation, which contains four partitions, \textit{i.e.}, train-360 (212 h), train-100 (58 h), dev (11 h) and test (11 h), and in this work we only use the train-360 partition for training.
The mixtures are created by randomly mixing utterances of two different speakers from LibriSpeech~\cite{panayotov2015librispeech}.
The noisy version is created by adding noise samples from WHAM!~\cite{wichern2019wham} dataset, which contains 82 hours of noise data recorded in coffee shops, restaurants and bars.
The resulting SNRs are normally distributed with a mean of -2 dB and a standard deviation of 3.6 dB.

Libri3Mix follows the same structure as Libri2Mix, \textit{i.e.}, train-360 (146 h), train-100 (40 h), dev (11 h) and test (11 h), where we only use the train-360 partition for training in this work.
The mixtures are created with three different speakers from LibriSpeech, and the noisy version is created similar to that of Libri2Mix.

\vspace{-0.3cm}
\subsection{Experimental Setup}
\label{ssec:exp_setup}

\vspace{-0.1cm}
\subsubsection{Network Configurations}
\label{sssec:network_configs}
\vspace{-0.15cm}
Following prior works~\cite{subakan2021attention, subakan2022using}, we employ 256 filters for the convolution layer in encoder, with kernel size of 16 and stride of 8, and decoder uses the same kernel size and stride as encoder.

The SE and SS networks in our system share the same architecture, which process chunks of size $K=250$ with 50\% overlap.
In SE network, the sequence modeling block contains 2 DPRNN blocks~\cite{luo2020dual} using BLSTM~\cite{hochreiter1997lstm} with 256 units in each direction, or 2 SepFormer blocks~\cite{subakan2021attention} with 1 Transformer~\cite{vaswani2017attention} layer in both IntraT and InterT.
In SS network, the sequence modeling block contains 6 DPRNN blocks using BLSTM with 256 units in each direction, or 2 SepFormer blocks with 8 Transformer layers in IntraT and InterT, with the attention heads/feed-forward dimension set to 8/1024.
 
\vspace{-0.35cm}
\subsubsection{Training Details}
\label{sssec:train_details}
\vspace{-0.15cm}
We train 200 epochs for all models, where Adam algorithm~\cite{kingma2014adam} is used for optimization, with initial learning rate of $1.5e^{-4}$.
After 85 epochs with DPRNN (5 epochs with SepFormer), the learning rate is halved if there is no improvement of validation performance for 5 consecutive epochs.
The batch size is set to 1.
Gradient clipping is used to limit the $L_2$ norm of gradients to 5.
No dynamic mixing~\cite{zeghidour2021wavesplit} strategy is used for data augmentation.
The weighting parameter $\lambda_\text{SE}$ is set to 0.1.
All hyper-parameters are tuned on validation set.

\begin{table}[t]
    \centering
    \vspace{-0.15cm}
    \caption{Comparison with the state-of-the-arts on Libri2Mix-noisy dataset.
    ``DPRNN'' or ``SepFormer'' in brackets denotes the backbone of sequence modeling block.
    * denotes self-reproduced results.}
    \vspace{0.15cm}
    \label{table1}
    \resizebox{0.47\textwidth}{!}{
    \begin{tabular}{p{10em}|c|c|c}
        \toprule
        Method & SI-SNRi (dB) & SDRi (dB) & \# Params \\
        \midrule
        ConvTasNet~\cite{luo2019conv} & 12.0 & 12.4 & 5.1 M \\
        Dual-Path RNN*~\cite{luo2020dual} & 14.2 & 14.7 & 14.6 M \\
        SepFormer~\cite{subakan2021attention} & 14.9 & 15.4 & 25.7 M \\
        Wavesplit~\cite{zeghidour2021wavesplit} & 15.1 & 15.8 & 29.0 M \\
        \midrule
        Ours (DPRNN) & 15.4 & 16.0 & 19.7 M \\
        Ours (SepFormer) & \textbf{16.0} & \textbf{16.5} & 29.2 M \\
        \bottomrule
    \end{tabular}}
    \vspace{-0.48cm}
\end{table}

\begin{table}[t]
    \centering
    \caption{Comparison with the state-of-the-arts on Libri3Mix-noisy dataset.
    * denotes self-reproduced results.}
    \vspace{0.15cm}
    \label{table2}
    \resizebox{0.47\textwidth}{!}{
    \begin{tabular}{p{10em}|c|c|c}
        \toprule
        Method & SI-SNRi (dB) & SDRi (dB) & \# Params \\
        \midrule
        ConvTasNet~\cite{luo2019conv} & 10.4 & 10.9 & 5.1 M \\
        Wavesplit~\cite{zeghidour2021wavesplit} & 13.1 & 13.8 & 29.0 M \\
        Dual-Path RNN*~\cite{luo2020dual} & 14.0 & 14.4 & 14.7 M \\
        SepFormer~\cite{subakan2021attention} & 14.3 & 14.8 & 25.7 M \\
        \midrule
        Ours (DPRNN) & 15.4 & 15.9 & 19.7 M \\
        Ours (SepFormer) & \textbf{15.8} & \textbf{16.4} & 29.2 M \\
        \bottomrule
    \end{tabular}}
    \vspace{-0.48cm}
\end{table}

\vspace{-0.28cm}
\subsection{Results}
\label{ssec:results}
\vspace{-0.1cm}

\subsubsection{Comparison with the State-of-the-Arts}
\label{sssec:compare_with_sota}
\vspace{-0.15cm}

\begin{table*}[t]
    \centering
    \vspace{-0.2cm}
    \caption{Effect of unified network and gradient modulation in our proposed approach with Libri2Mix-noisy dataset. 
    ``\# DP / SF'' denotes the number of DPRNN or SepFormer blocks.
    ``\# RNN / T'' denotes the number of RNN or Transformer layers in each Intra-/Inter-RNN or Intra-/Inter-Transformer.
    ``GM'' denotes whether use gradient modulation.
    ``\# Params'' denotes the total number of model parameters.}
    \vspace{0.15cm}
    \label{table3}
    \resizebox{0.95\textwidth}{!}{
    \begin{tabular}{l|cc|cc|c|c|c|c|c}
        \toprule
        \multirow{2}{*}{Method} & \multicolumn{2}{c|}{SE Network} & \multicolumn{2}{c|}{SS Network} & \multirow{2}{*}{$\lambda_\text{SE}$} &
        \multirow{2}{*}{GM} & \multirow{2}{*}{SI-SNRi (dB)} & \multirow{2}{*}{SDRi (dB)} & \multirow{2}{*}{\# Params}\\ \cline{2-5}
        & \# DP / SF & \# RNN / T & \# DP / SF & \# RNN / T & & & & & \\
        \midrule
        Dual-Path RNN~\cite{luo2020dual} & - & - & 6 & 1 & - & - & 14.2 & 14.7 & 14.6 M \\
        \midrule
        \multirow{4}{*}{Ours (DPRNN)} & 2 & 1 & 4 & 1 & 0 & \textcolor{lightgray}{\xmark} & 14.5 & 15.0 & 14.9 M \\
        & 2 & 1 & 6 & 1 & 0 & \textcolor{lightgray}{\xmark} & 14.6 & 15.1 & 19.7 M \\
        & 2 & 1 & 6 & 1 & 0.1 & \textcolor{lightgray}{\xmark} & 14.8 & 15.4 & 19.7 M \\
        & 2 & 1 & 6 & 1 & 0.1 & \cmark & \textbf{15.4} & \textbf{16.0} & 19.7 M \\
        \midrule
        SepFormer~\cite{subakan2021attention} & - & - & 2 & 8 & - & - & 14.9 & 15.4 & 25.7 M \\
        \midrule
        \multirow{4}{*}{Ours (SepFormer)} & 2 & 1 & 2 & 7 & 0 & \textcolor{lightgray}{\xmark} & 15.2 & 15.7 & 26.0 M \\
        & 2 & 1 & 2 & 8 & 0 & \textcolor{lightgray}{\xmark} & 15.3 & 15.7 & 29.2 M \\
        & 2 & 1 & 2 & 8 & 0.1 & \textcolor{lightgray}{\xmark} & 15.5 & 16.0 & 29.2 M \\
        & 2 & 1 & 2 & 8 & 0.1 & \cmark & \textbf{16.0} & \textbf{16.5} & 29.2 M \\
        \bottomrule
    \end{tabular}}
    \vspace{-0.45cm}
\end{table*}

Table~\ref{table1} compares our proposed approach with the best results of prior works on Libri2Mix-noisy dataset.
Our best system achieves the state-of-the-art with a SI-SNR improvement (SI-SNRi) of 16.0 dB and a Signal-to-Distortion Ratio improvement (SDRi) of 16.5 dB.
Our system with DPRNN and SepFormer backbones significantly outperform corresponding baselines (14.2 dB$\rightarrow$15.4 dB, 14.9 dB$\rightarrow$16.0 dB), while only cost a bit more model parameters.

Table~\ref{table2} compares our approach with prior works on Libri3Mix-noisy dataset, where our best system achieves the state-of-the-art with SI-SNRi result of 15.8 dB and SDRi of 16.4 dB.

As a result, our approach shows superior performance under noisy conditions, for separation of both two and three speakers.

\vspace{-0.3cm}
\subsubsection{Effect of Unified Network}
\label{sssec:effect_unified_net}
\vspace{-0.1cm}

We study the effect of unified network using Libri2Mix-noisy dataset in Table~\ref{table3}.
Compared to DPRNN baseline, our unified network with 2 DPRNN blocks in SE network and 4 blocks in SS network achieves better SI-SNRi result (14.2 dB$\rightarrow$14.5 dB) while without extra model parameters, indicating that front-end SE can benefit the downstream SS task.
Further increasing the DPRNN blocks in SS network brings more improvements (14.5 dB$\rightarrow$14.6 dB).
Based on this, adding SE loss for multi-task learning achieves higher SI-SNRi (14.6 dB$\rightarrow$14.8 dB), which benefits from the supervision information in clean mixture.
Similar improvements are observed on SepFormer backbone.

\vspace{-0.38cm}
\subsubsection{Effect of Gradient Modulation}
\label{sssec:effect_gm}
\vspace{-0.15cm}

\begin{figure}[t]
  \centering
  \includegraphics[width=0.98\columnwidth]{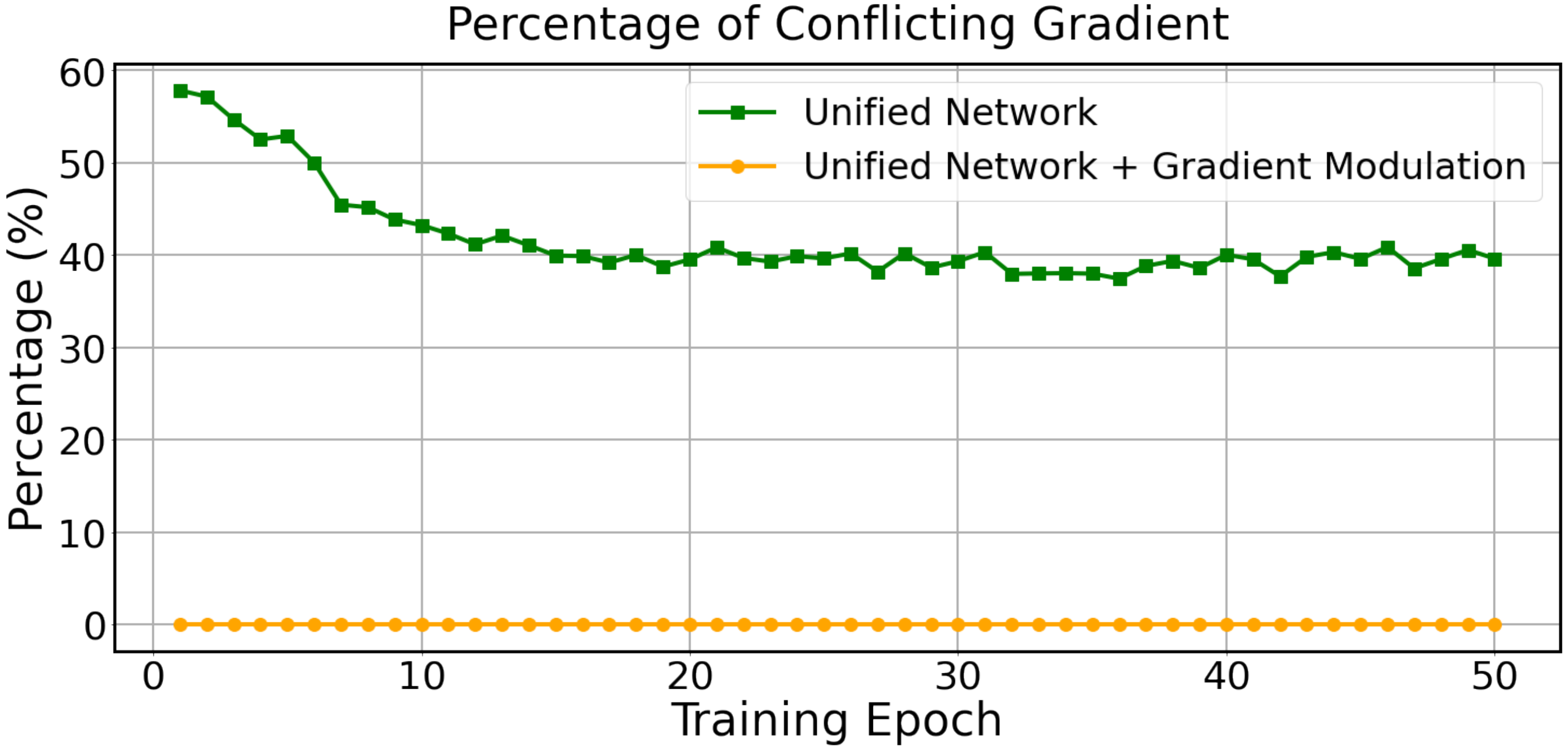}
  \vspace{-0.35cm}
  \caption{Percentage\% of conflicting gradient in all layers of SE network and encoder during training stage on Libri2Mix-noisy dataset.
  The percentage value of each epoch is obtained by averaging all the batches in it.
  We use SepFormer as backbone and present the first 50 epochs here (percentage value is stable in subsequent epochs).
  }\label{fig4}
  \vspace{-0.55cm}
\end{figure}

To illustrate the effect of proposed gradient modulation strategy, we present the percentage of conflicting gradient in all layers of SE network and encoder in Figure~\ref{fig4}.
We observe that the unified network with multi-task learning suffers a lot from gradient conflict (around 40\%).
In comparison, our gradient modulation strategy completely removes the conflicts, and thus alleviates the over-suppression problem as analyzed in Figure~\ref{fig5} and Section~\ref{sssec:visualization}.
As a result, we can observe significant improvements of the final SI-SNRi performance, \textit{i.e.}, 14.8 dB$\rightarrow$15.4 dB, 15.5 dB$\rightarrow$16.0 dB, as shown in Table~\ref{table3}.

\vspace{-0.36cm}
\subsubsection{Visualization of Mixture and Separated Speech}
\label{sssec:visualization}
\vspace{-0.15cm}

\begin{figure}[t]
  \centering
  \includegraphics[width=0.93\columnwidth]{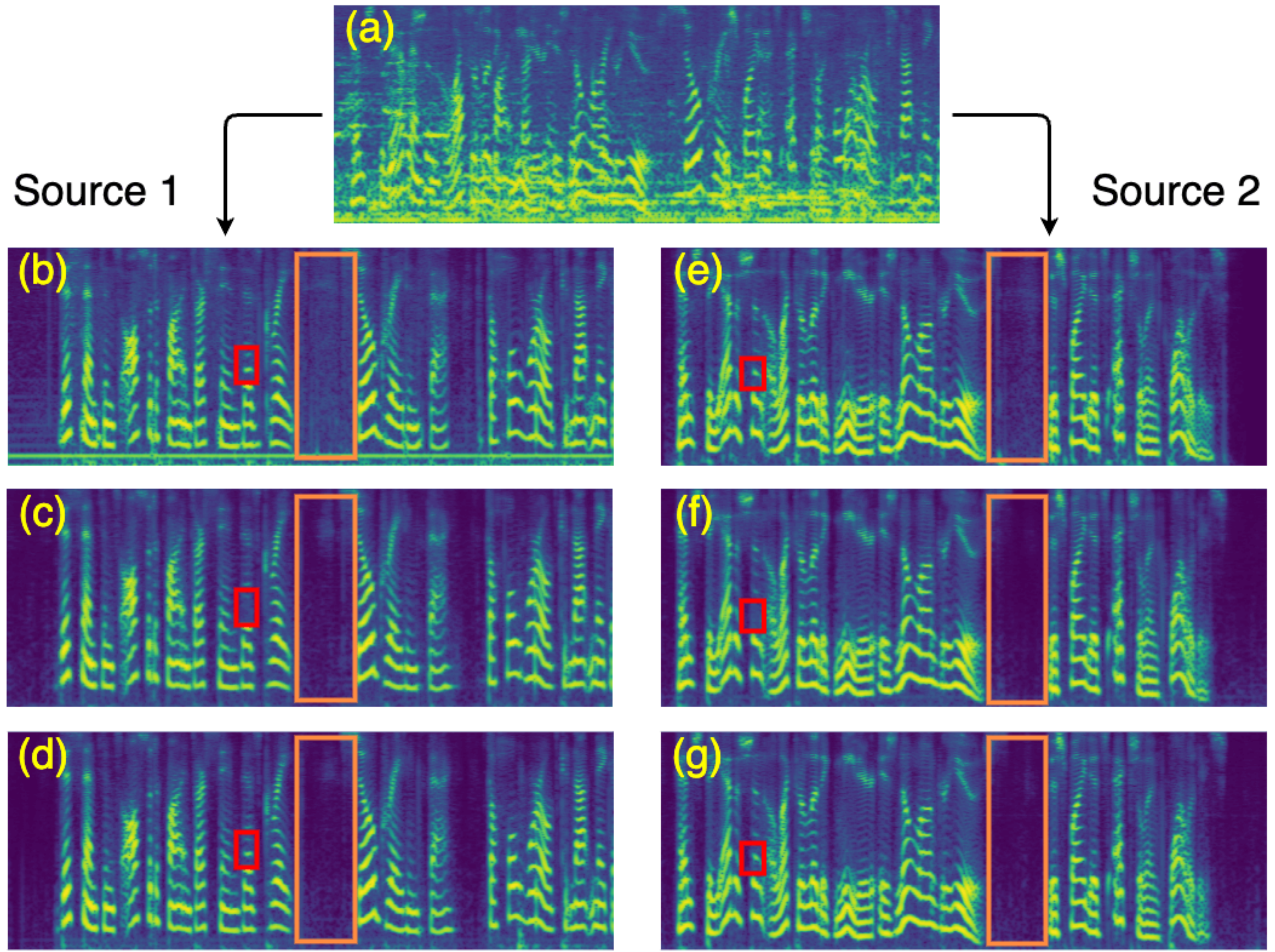}
  \vspace{-0.3cm}
  \caption{Spectrums of mixture and separated speech:
  (a) noisy mixture; separated source 1 in (b) SepFormer baseline, (c) our unified network, (d) our unified network + GM; source 2 in (e) SepFormer baseline, (f) our unified network, (g) our unified network + GM.}\label{fig5}
  \vspace{-0.4cm}
\end{figure}

To further show the overall effect of our approach, we visualize the spectrums of noisy mixture and separated speech using a test sample from Libri2Mix-noisy dataset, as presented in Figure~\ref{fig5}.
We first observe a lot of noise in the noisy mixture (a), which makes it difficult to separate each target source.
The SepFormer baseline separates the two sources from mixture as presented in (b) and (e), while we can still observe some noise in the separated speech (orange boxes).
In comparison, our unified network not only separates the two sources well, but also reduces the background noise, as shown in (c) and (f).
It indicates that speech enhancement with clean supervision information can effectively reduce noise for downstream speech separation.
However, we can also observe some loss of valid speaker information caused by SE, \textit{i.e.}, over-suppression (red boxes).
In comparison, our proposed gradient modulation strategy can alleviate this problem by harmonizing SE and SS tasks from optimization view, where some over-suppressed information is recovered as indicated by the red boxes in (d) and (g).
As a result, our proposed approach can effectively improve the noise-robustness of speech separation while avoid suppressing valid speaker information at the same time.

\vspace{-0.12cm}
\section{Conclusion}
\label{sec:conclusion}
\vspace{-0.2cm}
In this paper, we propose a novel network to unify speech enhancement and separation with gradient modulation for end-to-end noise-robust speech separation.
Specifically, we first build a unified network by combining speech enhancement and separation modules, with multi-task learning for optimization, where SE module is supervised by parallel clean mixture to reduce noise for downstream speech separation.
Furthermore, in order to avoid suppressing valid speaker information when reducing the noise, we propose a gradient modulation strategy to harmonize the SE and SS tasks from optimization view.
Experimental results demonstrate that our proposed approach improves the noise-robustness of speech separation model and achieves the state-of-the-art on large-scale benchmarks.

\vfill\pagebreak

\ninept
\bibliographystyle{IEEEbib}
% \bibliography{strings,refs}

\begin{thebibliography}{10}

\bibitem{luo2019conv}
Yi~Luo and Nima Mesgarani,
\newblock ``Conv-tasnet: Surpassing ideal time--frequency magnitude masking for
  speech separation,''
\newblock {\em IEEE/ACM transactions on audio, speech, and language
  processing}, vol. 27, no. 8, pp. 1256--1266, 2019.

\bibitem{luo2020dual}
Yi~Luo, Zhuo Chen, and Takuya Yoshioka,
\newblock ``Dual-path rnn: efficient long sequence modeling for time-domain
  single-channel speech separation,''
\newblock in {\em ICASSP 2020-2020 IEEE International Conference on Acoustics,
  Speech and Signal Processing (ICASSP)}. IEEE, 2020, pp. 46--50.

\bibitem{subakan2021attention}
Cem Subakan, Mirco Ravanelli, Samuele Cornell, Mirko Bronzi, and Jianyuan
  Zhong,
\newblock ``Attention is all you need in speech separation,''
\newblock in {\em ICASSP 2021-2021 IEEE International Conference on Acoustics,
  Speech and Signal Processing (ICASSP)}. IEEE, 2021, pp. 21--25.

\bibitem{subakan2022using}
Cem Subakan, Mirco Ravanelli, Samuele Cornell, Fran{\c{c}}ois Grondin, and
  Mirko Bronzi,
\newblock ``On using transformers for speech-separation,''
\newblock {\em arXiv preprint arXiv:2202.02884}, 2022.

\bibitem{zeghidour2021wavesplit}
Neil Zeghidour and David Grangier,
\newblock ``Wavesplit: End-to-end speech separation by speaker clustering,''
\newblock {\em IEEE/ACM Transactions on Audio, Speech, and Language
  Processing}, vol. 29, pp. 2840--2849, 2021.

\bibitem{vaswani2017attention}
Ashish Vaswani, Noam Shazeer, Niki Parmar, Jakob Uszkoreit, Llion Jones,
  Aidan~N Gomez, {\L}ukasz Kaiser, and Illia Polosukhin,
\newblock ``Attention is all you need,''
\newblock in {\em Advances in neural information processing systems}, 2017, pp.
  5998--6008.

\bibitem{chen2016noise}
Jitong Chen, Yuxuan Wang, and DeLiang Wang,
\newblock ``Noise perturbation for supervised speech separation,''
\newblock {\em Speech communication}, vol. 78, pp. 1--10, 2016.

\bibitem{wang2020complex}
Zhong-Qiu Wang, Peidong Wang, and DeLiang Wang,
\newblock ``Complex spectral mapping for single-and multi-channel speech
  enhancement and robust asr,''
\newblock {\em IEEE/ACM transactions on audio, speech, and language
  processing}, vol. 28, pp. 1778--1787, 2020.

\bibitem{liu2021ustc}
Dan Liu, Mengge Du, Xiaoxi Li, Yuchen Hu, and Lirong Dai,
\newblock ``The ustc-nelslip systems for simultaneous speech translation task
  at iwslt 2021,''
\newblock {\em arXiv preprint arXiv:2107.00279}, 2021.

\bibitem{chen2022self}
Chen Chen, Yuchen Hu, Nana Hou, Xiaofeng Qi, Heqing Zou, and Eng~Siong Chng,
\newblock ``Self-critical sequence training for automatic speech recognition,''
\newblock in {\em ICASSP 2022-2022 IEEE International Conference on Acoustics,
  Speech and Signal Processing (ICASSP)}. IEEE, 2022, pp. 3688--3692.

\bibitem{chen2022noise}
Chen Chen, Nana Hou, Yuchen Hu, Shashank Shirol, and Eng~Siong Chng,
\newblock ``Noise-robust speech recognition with 10 minutes unparalleled
  in-domain data,''
\newblock in {\em ICASSP 2022-2022 IEEE International Conference on Acoustics,
  Speech and Signal Processing (ICASSP)}. IEEE, 2022, pp. 4298--4302.

\bibitem{zhu2022noise}
Qiu-Shi Zhu, Jie Zhang, Zi-Qiang Zhang, Ming-Hui Wu, Xin Fang, and Li-Rong Dai,
\newblock ``A noise-robust self-supervised pre-training model based speech
  representation learning for automatic speech recognition,''
\newblock in {\em ICASSP 2022-2022 IEEE International Conference on Acoustics,
  Speech and Signal Processing (ICASSP)}. IEEE, 2022, pp. 3174--3178.

\bibitem{zhu2022joint}
Qiu-Shi Zhu, Jie Zhang, Zi-Qiang Zhang, and Li-Rong Dai,
\newblock ``Joint training of speech enhancement and self-supervised model for
  noise-robust asr,''
\newblock {\em arXiv preprint arXiv:2205.13293}, 2022.

\bibitem{zhu2022robust}
Qiu-Shi Zhu, Long Zhou, Jie Zhang, Shu-Jie Liu, Yu-Chen Hu, and Li-Rong Dai,
\newblock ``Robust data2vec: Noise-robust speech representation learning for
  asr by combining regression and improved contrastive learning,''
\newblock {\em arXiv preprint arXiv:2210.15324}, 2022.

\bibitem{pandey2021dual}
Ashutosh Pandey, Chunxi Liu, Yun Wang, and Yatharth Saraf,
\newblock ``Dual application of speech enhancement for automatic speech
  recognition,''
\newblock in {\em 2021 IEEE Spoken Language Technology Workshop (SLT)}. IEEE,
  2021.

\bibitem{hu2022interactive}
Yuchen Hu, Nana Hou, Chen Chen, and Eng~Siong Chng,
\newblock ``Interactive feature fusion for end-to-end noise-robust speech
  recognition,''
\newblock in {\em ICASSP 2022-2022 IEEE International Conference on Acoustics,
  Speech and Signal Processing (ICASSP)}. IEEE, 2022, pp. 6292--6296.

\bibitem{hu2022dual}
Yuchen Hu, Nana Hou, Chen Chen, and Eng~Siong Chng,
\newblock ``Dual-path style learning for end-to-end noise-robust speech
  recognition,''
\newblock {\em arXiv preprint arXiv:2203.14838}, 2022.

\bibitem{he2015delving}
Kaiming He, Xiangyu Zhang, Shaoqing Ren, and Jian Sun,
\newblock ``Delving deep into rectifiers: Surpassing human-level performance on
  imagenet classification,''
\newblock in {\em Proceedings of the IEEE International Conference on Computer
  Vision (ICCV)}, December 2015.

\bibitem{glorot2011deep}
Xavier Glorot, Antoine Bordes, and Yoshua Bengio,
\newblock ``Deep sparse rectifier neural networks,''
\newblock in {\em Proceedings of the fourteenth international conference on
  artificial intelligence and statistics}. JMLR Workshop and Conference
  Proceedings, 2011, pp. 315--323.

\bibitem{le2019sdr}
Jonathan Le~Roux, Scott Wisdom, Hakan Erdogan, and John~R Hershey,
\newblock ``Sdr--half-baked or well done?,''
\newblock in {\em ICASSP 2019-2019 IEEE International Conference on Acoustics,
  Speech and Signal Processing (ICASSP)}. IEEE, 2019, pp. 626--630.

\bibitem{kolbaek2017multitalker}
Morten Kolb{\ae}k, Dong Yu, Zheng-Hua Tan, and Jesper Jensen,
\newblock ``Multitalker speech separation with utterance-level permutation
  invariant training of deep recurrent neural networks,''
\newblock {\em IEEE/ACM Transactions on Audio, Speech, and Language
  Processing}, vol. 25, no. 10, pp. 1901--1913, 2017.

\bibitem{cosentino2020librimix}
Joris Cosentino, Manuel Pariente, Samuele Cornell, Antoine Deleforge, and
  Emmanuel Vincent,
\newblock ``Librimix: An open-source dataset for generalizable speech
  separation,''
\newblock {\em arXiv preprint arXiv:2005.11262}, 2020.

\bibitem{panayotov2015librispeech}
Vassil Panayotov, Guoguo Chen, Daniel Povey, and Sanjeev Khudanpur,
\newblock ``Librispeech: an asr corpus based on public domain audio books,''
\newblock in {\em 2015 IEEE international conference on acoustics, speech and
  signal processing (ICASSP)}. IEEE, 2015, pp. 5206--5210.

\bibitem{wichern2019wham}
Gordon Wichern, Joe Antognini, Michael Flynn, Licheng~Richard Zhu, Emmett
  McQuinn, Dwight Crow, Ethan Manilow, and Jonathan Le~Roux,
\newblock ``Wham!: Extending speech separation to noisy environments,''
\newblock {\em Proc. Interspeech 2019}, pp. 1368--1372, 2019.

\bibitem{hochreiter1997lstm}
Sepp Hochreiter and Jürgen Schmidhuber,
\newblock ``Long short-term memory,''
\newblock {\em Neural Computation}, vol. 9, no. 8, pp. 1735--1780, 1997.

\bibitem{kingma2014adam}
Diederik~P. Kingma and Jimmy Ba,
\newblock ``Adam: A method for stochastic optimization,''
\newblock {\em arXiv preprint arXiv:1412.6980}, 2014.

\end{thebibliography}

\end{document}